\documentclass[journal,onecolumn,draftcls,12pt]{IEEEtran}
\usepackage[T1]{fontenc}

\makeatletter
\def\endthebibliography{%
	\def\@noitemerr{\@latex@warning{Empty `thebibliography' environment}}%
	\endlist
}
\makeatother

\IEEEoverridecommandlockouts

\usepackage{tikz}
    \usetikzlibrary{shapes.arrows}

\usepackage{pgfplots}
\usepackage{adjustbox}

\usepackage{gincltex}
\usepgfplotslibrary{fillbetween}
\usepgfplotslibrary{colorbrewer}
\usetikzlibrary{plotmarks}
\usetikzlibrary{patterns}

\usepackage{cite}
\usepackage{amsmath,amssymb,amsfonts}
\usepackage{algorithm}
\usepackage{algorithmic}
\usepackage{etoolbox}  
\usepackage{bbm}
\usepackage{amsmath}
\usepackage{amssymb}
\usepackage{amsthm}
\usepackage{bm}
\usepackage{bbm}
\usepackage{multicol, latexsym, amsmath, amssymb}
\usepackage{blindtext}
\usepackage{caption}
\usepackage{tabu}
\usepackage{booktabs}
\usepackage{anysize} 
\usepackage{hyperref}
\usepackage{subfig}
\usepackage{glossaries}

\newacronym{iot}{IoT}{Internet of Things}
\newacronym{nbiot}{NB-IoT}{narrowband Internet of Things}
\newacronym{nr}{NR}{New Radio}
\newacronym{ntn}{NTN}{Non-Terrestrial Networks}
\newacronym{leo}{LEO}{Low-Earth Orbit}
\newacronym{meo}{MEO}{Medium-Earth Orbit}
\newacronym{geo}{GEO}{Geo-synchronous Earth Orbit}
\newacronym{cn}{CN}{Core Network}
\newacronym{isl}{ISL}{Inter-Satellite Link}
\newacronym{gnb}{gNB}{Next Generation NodeB}
\newacronym{aoi}{AoI}{Age of Information}
\newacronym{ue}{UE}{User Equipment}
\newacronym{pdcch}{PDCCH}{physical downlink control channel}
\newacronym{prach}{PRACH}{physical random access channel}
\newacronym{nprach}{NPRACH}{narrowband \gls{prach}}
\newacronym{npdcch}{NPDCCH}{narrowband \gls{pdcch}}
\newacronym{pusch}{PUSCH}{physical uplink shared channel}
\newacronym{pdsch}{PDSCH}{physical downlink shared channel}
\newacronym{ra}{RA}{random access}
\newacronym{rao}{RAO}{random access opportunity}
\newacronym{dl}{DL}{downlink}
\newacronym{ul}{UL}{uplink}
\newacronym{gs}{GS}{Ground Station}
\newacronym{5gc}{5GC}{5G Core Network}
\newacronym{upf}{UPF}{User Plane Function}
\newacronym{ngso}{NGSO}{Non-geostationary orbit}
\newacronym{rtt}{RTT}{round-trip time}
\newacronym{tmtc}{TMTC}{telemetry and telecontrol}
\newacronym{pmf}{pmf}{probability mass function}
\newacronym{qos}{QoS}{Quality of Service}
\newacronym{fcfs}{FCFS}{First Come First Serve}
\newacronym{sinr}{SINR}{signal-to-interference-plus-noise ratio}
\newacronym{harq}{HARQ}{hybrid automatic repeat request}
\newacronym{embb}{eMBB}{Enhanced Mobile Broad-Band}
\newacronym{mmtc}{mMTC}{massive machine-type communication}
\newacronym{urllc}{URLLC}{ultra-reliable low-latency communication}
\newacronym{rrc}{RRC}{Radio Resource Control}

\makeatletter
\patchcmd{\algorithmic}{\addtolength{\ALC@tlm}{\leftmargin} }{\addtolength{\ALC@tlm}{\leftmargin}}{}{}
\makeatother

\makeatletter
\newcommand\fs@betterruled{%
	\def\@fs@cfont{\bfseries}\let\@fs@capt\floatc@ruled
	\def\@fs@pre{\vspace*{5pt}\hrule height.8pt depth0pt \kern2pt}%
	\def\@fs@post{\kern2pt\hrule\relax}%
	\def\@fs@mid{\kern2pt\hrule\kern2pt}%
	\let\@fs@iftopcapt\iftrue}
\floatstyle{betterruled}
\restylefloat{algorithm}
\makeatother

\usepackage{filecontents}
\usepackage{graphicx}
\usepackage{textcomp}
\usepackage{xcolor}
\usepackage{authblk}
\def\BibTeX{{\rm B\kern-.05em{\sc i\kern-.025em b}\kern-.08em
		T\kern-.1667em\lower.7ex\hbox{E}\kern-.125emX}}

\pgfplotsset{compat=1.15}
\begin{document}
\title{5G satellite networks for IoT: offloading and backhauling}
\author{Beatriz Soret,
Israel Leyva-Mayorga,
Stefano Cioni,
        and~Petar~Popovski
\thanks{B. Soret (corresponding author, email: bsa@es.aau.dk), I. Leyva-Mayorga, and P. Popovski are with the Department of Electronic Systems, Aalborg University, 9220 Aalborg, Denmark. S. Cioni is with European Space Agency - ESTEC, Noordwijk, The Netherlands.}}

\maketitle

\begin{abstract}
One of the main drivers of 5G cellular networks is provision of connectivity service for various Internet of Things (IoT) devices. 
Considering the potential volume of IoT devices at a global scale, 
the next leap is to integrate Non-Terrestrial Networks (NTN) into 5G terrestrial systems, thereby extend the coverage and complement the terrestrial service. This paper focuses on the use of Low-Earth Orbit (LEO) satellite constellations for two specific purposes: offloading and backhauling. The former allows offloading IoT traffic from a congested terrestrial network, usually in a very dense area. In the latter application, the constellation provides a multi-hop backhaul that connects a remote terrestrial gNB to the 5G core network. After providing an overview of the status of the 3GPP standardization process, we model and analyze the user data performance, specifically in the uplink access and the satellite multi-hop constellation path. The evaluation of the collisions, the delay and the Age of Information, and the comparison of the terrestrial and the satellite access networks provide useful insights to understand the potential of LEO constellations for offloading and backhauling of IoT traffic.
\end{abstract}

\IEEEpeerreviewmaketitle
\section{Introduction}

\gls{leo} constellations, deployed between 500 and 2000 km over the Earth surface, represent a promising technology to complement 5G terrestrial networks in the \gls{iot} area \cite{Leyva-Mayorga2020}\cite{Guidotti2019}. Unlike geostationary orbits, \gls{leo} satellites have a small ground coverage and move fast with respect to the Earth's surface. \gls{leo} constellations are typically organized in orbital planes, where satellites follow the same trajectory, one after the other. To provide continuous coverage with low orbits, a densely populated constellation is required to ensure that, at least, one satellite is always available to any ground device. The high number of spacecrafts to be deployed calls for the use of \emph{low-cost} small satellite technology~\cite{Saeed2020} to reduce the total cost, but this imposes strong power constraints. With limited power, the satellites in the orbital plane must stay close to each other in order to support an \gls{isl}. Two distant points on the Earth's surface are typically connected through a multi-hop connection and some of these hops take place over \gls{isl}s  \cite{Leyva-Mayorga2020}. Connectivity becomes more challenging when communication between different orbital planes is required, due to the relative speed and time-varying inter-satellite distance. 

3GPP has recently started the work to define the role of satellite communications, also known as \gls{ntn}~\cite{Wang2020}\cite{3GPPTR38.811}, in future releases of 5G \gls{nr}. This includes \gls{leo} satellite constellations, also referred to as \gls{ngso} satellites. In particular, \gls{leo}  constellations  are  envisioned to  dramatically extend cellular coverage,  serve as a global backbone, and offload  the cellular base stations in congested areas~\cite{3GPPTR22.822, TR38.913, 3GPPTR38.811, 3GPPTR38.821}. One of the envisioned applications is \gls{ntn} \gls{iot}.
\gls{iot} communications have requirements and constraints that are significantly different from the broadband connections. Indeed, in order to complement the 5G \gls{nr} \gls{embb} services, 3GPP has defined two additional connectivity types: \gls{mmtc} and \gls{urllc}. They are both characterized by the transmission of small packets, but in \gls{mmtc} the huge amount of \gls{iot} devices competing for the shared resources is the most critical requirement, whereas in \gls{urllc} the main challenge is the need to reliably deliver data in a very short time.
\gls{mmtc} and \gls{urllc} were originally defined for terrestrial networks and their features and requirements should be reconsidered in the physical context of \gls{ntn}. Naturally, the most stringent \gls{urllc} requirement of 1 ms latency is infeasible with a satellite connection. In this regard, the \gls{rtt} latency has been redefined to 50 ms for \gls{leo} satellite systems~\cite{TR38.913}. Nevertheless, other critical \gls{iot} requirements such as reliability and connection density remain unchanged in 5G \gls{nr} Release 15. The footprint and the mobility of a satellite working as a base station (\gls{gnb} in 5G \gls{nr} terminology) are also fundamentally different from the fixed terrestrial infrastructure.  

Furthermore, in many practical \gls{iot} scenarios, the use of the network latency as the requirement for the timing is constricting. To elaborate, there are applications where a source generates updates that are transmitted through a communication network, such as tracking containers in logistics, data from ships/vessels using VHF data exchange system (VDES) \cite{ITU2371} \cite{Lazaro2019} or Authentication Identification System (AIS) \cite{ITU_AIS} or Automatic Dependent Surveillance - Broadcast (ADS-B) \cite{ADSB} system in airplanes. Another group of applications interested in the timeliness of the information are sensor networks, where real-time status updates of a process are transmitted from a generating point to a remote destination. For all these \gls{iot} cases, the end receiver is interested in the freshest update, rather than the per-packet delay, and the classical definition of network delay has been recently augmented by the concept of \gls{aoi} \cite{Kaul2012}. 


Motivated by the current interest in integrating \gls{leo} constellations in the 5G specification, we investigate their potential to support \gls{iot} data traffic and the achievable performance in two exemplary use cases. In the first scenario, a \gls{leo} constellation offloads \gls{iot} traffic from a congested terrestrial network, usually in a particularly densely deployed area, by offering an alternative interface for the devices' traffic. The focus in this case is on the diversity offered by the additional access taking into account the necessary time scaling of the access channels and parameters to adapt to the characteristics of \gls{leo} \gls{ntn}. The second scenario comprises a remote \gls{gnb}, typically in a rural area, that uses the \gls{leo} constellation as the backhaul to connect the terrestrial \gls{gnb} to the 5G core side where there is no cost‐effective terrestrial backhaul option. We investigate here the network performance when the space segment connects two distant points, for which not only the access introduces a delay but the \gls{isl} is necessary to route the information from origin to destination. While implementing the \gls{isl} is not mandatory, it provides a valuable feature to achieve a convenient trade-off between the density of deployment of the constellation and of satellite gateways on ground to ensure end-to-end communications links between the source and the destination.

The rest of the paper is organized as follows. In Section \ref{sec:5g} we discuss the status of \gls{ntn} \gls{iot} in 3GPP and the relevant architecture elements. Section \ref{sec:systemmodel} presents the system model, and  Sections \ref{sec:offloading} and \ref{sec:backhauling} analyze and evaluate the performance of the offloading and backhauling scenarios, respectively. Section \ref{sec:conclusions} has the conclusions and future work. 

\section{LEO constellations for 5G NTN IoT}
\label{sec:5g}
\subsection{LEO satellite constellations}
\gls{leo} satellites are deployed at altitudes between 500 and 2000 km. Unlike geostationary orbits, \gls{leo} satellites move fast with respect to the Earth's surface and have a small ground coverage: for instance, only 0.45 \% of the Earth's surface for a \gls{leo} satellite deployed at 600 km and with an elevation angle of 30 degrees. To ensure that any ground terminal is always covered by, at least, one satellite, a flying swarm of many satellites is required, usually organized in a \emph{constellation} with coordinated ground coverage~\cite{walker1971circular,Leyva-Mayorga2020}. 

A feeder link connects each satellite in the constellation with the \gls{gs} (or satellite gateway), but the continuity and availability of this link is limited by the number of \gls{gs}s and the satellite speed. The feeder link is vital for the \gls{tmtc} functions of a spacecraft, used for the spacecraft control and command. Moreover, the satellites can have direct connection with \gls{ue}s deployed on the Earth's surface. The constellation is organized in several orbital planes, where satellites are evenly distributed and inter-connected using the intra-plane \gls{isl}. They can also communicate with satellites in other planes using the inter-plane \gls{isl}. 

If a sufficient number of satellites are deployed, the constellation may offer a continuous service.

\subsection{3GPP NTN}
3GPP is working in the integration of \gls{ntn} in future releases of 5G \gls{nr}~\cite{3GPPTR38.811, 3GPPTR22.822, 3GPPTR38.821}. This encompasses \gls{leo}, \gls{meo}, and \gls{geo} satellites, but also air-borne vehicles, such as High Altitude Platforms (HAPs) operating typically at altitudes between $8$ and $50$~km.

Three use cases are considered in \gls{ntn}: (1) \emph{Service continuity}, to ensure a continuous coverage to mobile ground terminals that have been previously granted access to 5G services, such as terrestrial vehicles, ships, and airborne platforms; (2) \emph{Service ubiquity}, to provide 5G access in areas without terrestrial coverage, including areas where the terrestrial coverage is interrupted by a natural disaster, such as earthquake or flood; (3) \emph{Service scalability}, to give support to the terrestrial infrastructure in massive multicasting (downlink) or \gls{iot} (uplink) applications, as in ultra-high definition TV and ultra-dense \gls{iot} deployments.

In the context of the \gls{ntn} work item, the goal of 3GPP is to ensure an end-to-end standard in the Release 17 timeframe -- the third phase of 5G -- originally scheduled for 2021. In this first release, the work item baseline is the adaptation of \gls{embb} services via satellite networks, however a second feasibility study (or study item) for \gls{ntn} \gls{iot} was agreed in December 2019, with the goal of introducing both \gls{nbiot} and evolved MTC (eMTC) support from space. The study item will look at the modifications required to handle the large delay experienced in the satellite scenario and its impact in random-access procedures, the \gls{harq} operation and the \gls{rrc} procedures. The former is especially relevant for \gls{iot}. Another important area is the beam management and cell selection/re-selection \cite{RP19234}.

Two 5G satellite implementations are envisioned: transparent and regenerative payload~\cite{3GPPTR38.821}. In the first one, the satellites merely serve as relays toward the ground and, in the second one, satellites are fully or partially functional \glspl{gnb} that can perform, for example, encoding/decoding and routing. Hence, the regenerative payload implementation enables the use of the 5G logical interface between \glspl{gnb}, the so-called Xn, to connect distant \glspl{gnb} through the constellation. Moreover, 3GPP considers two options of multi-connectivity in \gls{ntn}, having the \gls{ue} connected to one satellite and one terrestrial network, or to two satellites~\cite{3GPPTR38.821}. In both cases, connectivity combining can occur for either the uplink or the downlink or both. 

\subsection{5G Core Network and Initial Access Procedure}
The 5G \gls{nr} network consists of \glspl{gnb} connected to the \gls{5gc}. \gls{5gc} presents several novelties as compared to LTE~\cite{3GPPTR23501}.  
It has a full control plane / user plane separation for better edge deployment and support of latency-sensitive services. On the data plane side, the \gls{upf} performs packet processing and transmission operations. The N3 and N6 connect the \gls{upf} to the \gls{gnb} and to the packet data network, respectively. A \gls{ue} can be connected to multiple \gls{upf}s simultaneously, e.g., one remote and one close to the edge for edge / low latency services. Another novelty is the split of the mobility-related tasks into two elements: (1) the 5G Core Access and Mobility Management Function (AMF) to handle the connection and mobility management tasks, for which it receives all connection and session related information from the UE (interfaces N1 and N2); (2) and the Session Management Function (SMF), responsible of all messages related to session management.

On the initial access procedure, the devices that attempt to communicate through a \gls{gnb}, either terrestrial or non-terrestrial, must first complete the \gls{ra} procedure. This allows them to switch from \gls{rrc} idle mode to \gls{rrc} connected mode. It shall be remembered that the current 5G \gls{mmtc} communication is based on the LTE technologies: LTE-M and \gls{nbiot}~\cite{Wang2017}. Throughout the rest of the paper and our performance evaluations, we will consider the RA procedure and parameters of \gls{nbiot}~\cite{3GPPTS36.321,3GPPTS36.331}, assuming that a similar protocol will be defined in the near future for 5G.

The standard 3GPP \gls{ra} procedure consists of four message exchange: preamble (\emph{Msg1}), uplink grant (\emph{Msg2}), connection request (\emph{Msg3}), and contention resolution (\emph{Msg4}). Out of these, \emph{Msg3} and \emph{Msg4} are scheduled transmissions where no contention takes place. Only after the completion of the \gls{ra} procedure, data can be transmitted in the uplink through the \gls{pusch} or in the downlink through the \gls{pdsch}. 

To reduce the latency and optimize the signaling, a two-step version is under discussion for Release 17~\cite[Section 7.2.1.1.2]{3GPPTR38.821}. This is achieved by transmitting the preamble (\emph{Msg1}) followed by the data transmission through the \gls{pusch}. Thence, the preamble serves as a temporary identifier for the \gls{ue} that sends the \gls{pusch} transmission, and the contention resolution message serves as an ACK. 

The preambles are orthogonal resources used to perform the \gls{ra} request (\emph{Msg1}). In 4G and 5G, Zadoff-Chu sequences are used, due to their good auto- and cross-correlation properties. However, they are difficult to generate and therefore preambles in \gls{nbiot} are simply unique single-tone and pseudo-random hopping sequences. These sequences are orthogonal and defined by the initial subcarrier. Hence, the number of available subcarriers in the \gls{nprach} $R\in\{12,24,36,48\}$ is equal to the number of available orthogonal preambles~\cite{3GPPTS36.331}. 

\begin{figure}[t]
    \centering
    \subfloat[]{\includegraphics{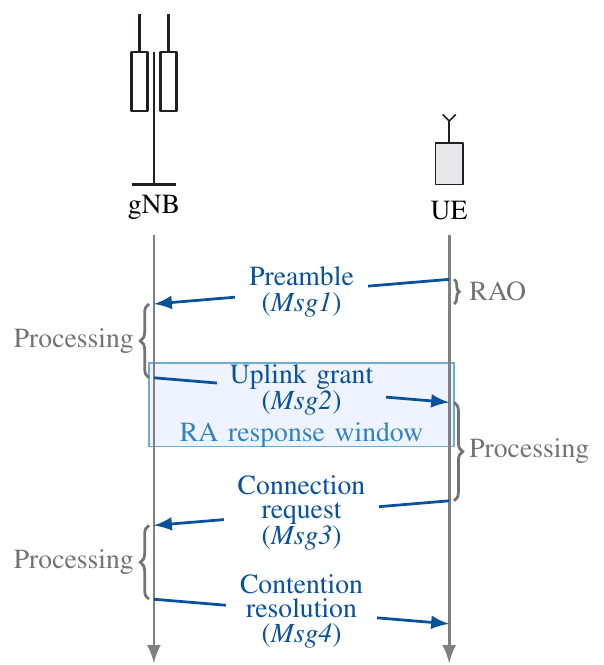}}\hfil
    \subfloat[]{\includegraphics{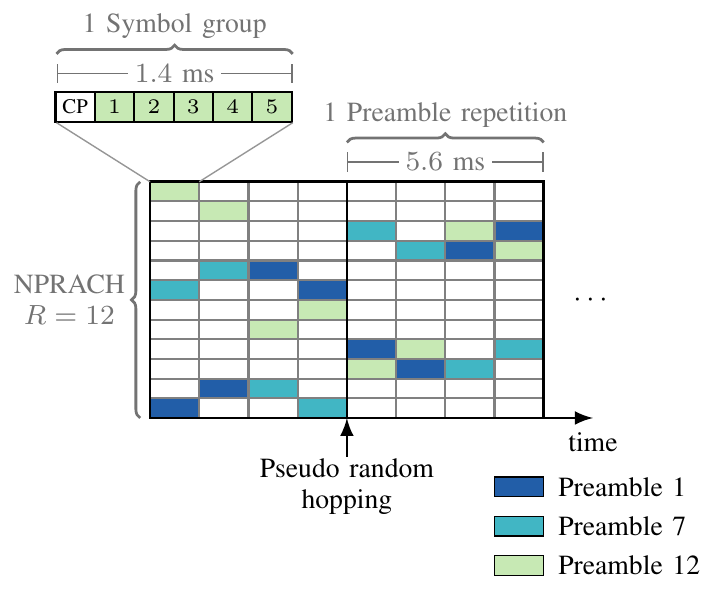}}
    \caption{Diagram of the (a) four-message handshake of the \gls{ra} procedure and (b) time and frequency resources in the \gls{nprach} for preamble transmission (with preamble format 0)~\cite{3GPPTS36.211,Lin2016} in \gls{nbiot}.}
    \label{fig:ra_diagram}
\end{figure}

The \gls{nprach} is scheduled to occur periodically in specific subframes; these are reserved for the RA requests and are commonly known as \glspl{rao}. To initiate the \gls{ra} procedure, the devices simply select the initial subcarrier randomly, generate the hopping sequence, and transmit it at the next available \gls{rao}. The orthogonality of preambles implies that multiple \glspl{ue} can access the \gls{gnb} in the same \gls{rao} if they select different preambles. Next, the grants are transmitted to the devices through the \gls{npdcch} within a predefined period known as the  \emph{\gls{ra} response window}. Figure~\ref{fig:ra_diagram} summarizes the \gls{ra} procedure.

Preambles with format 0 -- used throughout this paper -- consist of four contiguous symbol groups. Each of these symbol groups has a duration of $1.4$~ms, for a total preamble duration of $5.6$~ms~\cite[Section 10.1.6]{3GPPTS36.211}. Furthermore, \gls{nbiot} \glspl{ue} are placed in one of three possible coverage enhancement (CE) levels based on the quality of their wireless channel. Specifically, CE level $0$, $1$, and $2$ correspond to good, medium, and poor wireless conditions, respectively. Different number of repetitions (i.e., replicas) may be assigned to each CE level. These replicas are added coherently to enhance the \gls{sinr} of the transmissions and, hence, reduce erasure probability. In exchange, the energy consumption and the duration of the messages increase with the number of repetitions. For instance, let $\ell_\text{rep}$ be the number of repetitions for a specific CE level. Then, the preamble duration for the same CE level is  $5.6\,\ell_\text{rep}$~ms. The same applies to other messages, including the \gls{ra} response.

Naturally, scheduling the \gls{nprach} and \gls{npdcch} consumes resources that would otherwise be used for data transmission. Therefore, each  implementation must find an adequate balance between the amount of resources dedicated to \gls{nprach}, \gls{npdcch}, \gls{pusch}, and \gls{pdsch}. Hence, cells serving a large number of users that continuously generate and consume data, for example, in dense urban scenarios, may be forced to dedicate a low amount of resources to the \gls{nprach} and \gls{npdcch}. As a result, the period between \glspl{rao} may become large. 

\subsection{Offload and backhaul architectures}
 \begin{figure}
    \centering
    \includegraphics[width=6.5in]{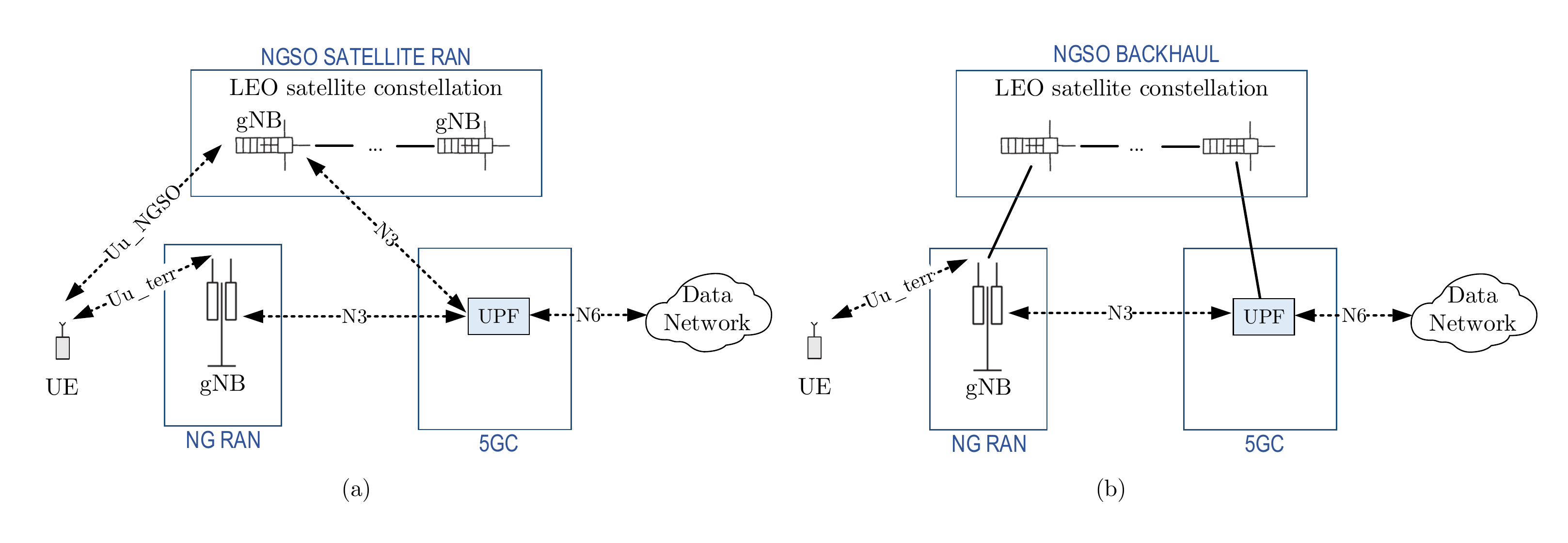}
    \caption{User plane with NGSO: (a) satellite offloading (b) satellite backhauling.} 
    \label{fig:backhauling_offloading}
\end{figure}

We consider a 5G terrestrial network consisting of \glspl{gnb} and a \gls{5gc} connected to the data network. There is also a \gls{leo} satellite constellation consisting of $N$ satellites organized in several orbital planes.  
The satellites in the constellation are regenerative payloads inter-connected via the \gls{isl}. 
The focus of the paper is on two main functionalities of a \gls{leo} constellation in a 5G system: to offload the terrestrial network and to serve as a backhaul  (see Figure \ref{fig:backhauling_offloading}).

\noindent \textbf{Offloading:} A \gls{leo} constellation can be used for data offloading when the terrestrial network is congested in very dense urban areas. Since we consider the regenerative satellite architecture, where a full \gls{gnb} is on-board of the satellite as payload, we consider the existence of dual-\glspl{ue} that can connect to the terrestrial or the satellite \gls{gnb}. The offloading policies should exploit the potential of the constellation to cover a wider area, but should also take into account the different \glspl{rtt} and channel configurations with respect to the terrestrial \glspl{gnb}. The satellite might download the traffic to the 5GC directly or use several other satellites in the constellation to reach the \gls{dl} connection to the terrestrial network or the final destination. Offloading traffic from the terrestrial system opens up the possibility of improving the reliability and latency performance, while supporting a larger number of users. Load balancing has long been studied as an approach to balance the workload across various \glspl{gnb}, e.g., in Heterogeneous Networks (HetNets) \cite{Andrews2014}. It can be used to optimize quantities like resource utilization, fairness, delays, or throughput.

\noindent \textbf{Bakchauling:} Backhaul for terrestrial mobile networks is critical to ensure speed and capacity as it relates to the transport of the packets from the distributed access network to the 5GC. Mobile operators have stayed away from satellite solutions for the backhaul, preferring fibre and and microwave solutions, mainly due to cost, latency and availability concerns. Nevertheless, the cost reduction of the New Space era and the use of low orbits places the satellite backhaul as a feasible solution to connect the terrestrial access network in remote areas where there is no cost-effective terrestrial option to reach the \gls{5gc}. Thus, the satellite system transparently carries the payload between the two communication extremes. Interestingly, the total latency through a dense \gls{leo} constellation is potentially similar or even lower than a terrestrial connection when the distance between the endpoints is large. This is due to a faster propagation speed of the electromagnetic waves in space than in fiber~\cite{Handley2018}. Furthermore, extensions to the 5GC functionalities to identify the UEs which are accessing via satellite 3GPP access are being considered. This is important for the integration of \gls{ntn} with \gls{qos} requirements, as it allows the use of advanced resource allocation and routing mechanisms to provide performance guarantees to different service types.

The option of a multilink backhauling, where the traffic between the 5GC and the RAN is transmitted through one or more satellite links and one or more terrestrial links, is also under discussion in 3GPP, but it is not considered in this paper. In that configuration, the traffic could be steered, switched or/and split over two links between the 5GC and the RAN. 

\section{System model} \label{sec:systemmodel}
Both cases, offloading and backhauling, are modeled with an uplink scenario with a potentially massive number of devices inside the coverage area of a terrestrial \gls{gnb} and of up to one space \gls{gnb} belonging to a \gls{leo} constellation. Specifically, we define $U$ as the total number of devices covered by the terrestrial \gls{gnb}; these are indexed by $u\in\{1,2,\dotsc,U\}$. Each \gls{iot} device generates status updates at a Poisson rate $\lambda_{u}$ updates per second, hence, the total update rate is $\lambda_{\text{tot}} = \sum_{u=1}^{U}\lambda_{u}$.

A fraction $\kappa$ of the \gls{iot} nodes attempt to access the terrestrial \gls{gnb} and the remaining $1-\kappa$ attempt to access the space \gls{gnb}. Therefore, the total update rates to the terrestrial and to the space \glspl{gnb} are $\lambda_\text{earth}=\kappa\lambda_{\text{tot}}$ and $\lambda_\text{space}=(1-\kappa)\lambda_{\text{tot}}$.

In the access, terrestrial or spatial, we model the \gls{nprach} for preamble transmission and \gls{npdcch} for \gls{ra} response. Due to the limited capacity of these channels, collisions, and wireless channel errors, only a fraction of the total rate will be relayed by the space and terrestrial \glspl{gnb}. Specifically, let $P_f^\text{earth}$ and $P_f^\text{space}$ be the probability of failure in the \gls{ra} procedure at the terrestrial and space \glspl{gnb}, respectively. Naturally, the probabilities of failure are a function of the capacity of the \gls{nprach} and \gls{npdcch} and of the update rates towards the \glspl{gnb}. 

Let $\mathbbm{1}_\text{5GC}$ be the indicator variable that takes the value of $1$ if the terrestrial \gls{gnb} has a terrestrial connection to the 5GC and $0$ otherwise. Hence, the rate that the terrestrial \gls{gnb} relays directly towards the 5GC is \begin{equation}
    \lambda_\text{5GC}^\text{earth}= \kappa \lambda_{\text{tot}}(1-P_f^\text{earth})\mathbbm{1}_\text{5GC}. \label{eq:eq_lambda_earth}
\end{equation} 

The rate that the space \gls{gnb} relays towards the 5GC is 
\begin{equation}
    \lambda^\text{space}_\text{5GC}=(1-\kappa)\lambda_{\text{tot}}(1-P_f^\text{space}) + \kappa \lambda_{\text{tot}}(1-P_f^\text{earth})(1-\mathbbm{1}_\text{5GC}).
\end{equation}

In our evaluations of the backhauling scenario all \glspl{ue} connect to the terrestrial gNB using the $\text{Uu}\_ \text{terr}$ interface and therefore $\kappa = 1$ and $\mathbbm{1}_\text{5GC} = 0$.


As mentioned before, we consider the \gls{ra} procedure and parameters of \gls{nbiot}, assuming that a similar protocol will be defined in the near future for 5G. In \gls{nbiot}, \Glspl{rao} occur once every $T_\text{rao}\in\{40,80,160,\dotsc,5120\}$~ms~\cite{3GPPTS36.331}.
At each \gls{rao}, the preamble transmission is modeled as a multichannel slotted ALOHA with $R$ available channels.  The number of \gls{ra} response messages per \gls{rao} are determined by the number of consecutive subframes that comprise the \emph{\gls{ra} response window} and the maximum number of \gls{ra} responses that can be sent per subframe in this window. The \gls{npdcch} has a duration of $0.5$~milliseconds and contains up to two \emph{control channel elements}. Therefore, up to four control channel elements can be scheduled in each subframe of duration $1$~millisecond~\cite{3GPPTS36.211}. 
Hence, we adopt the widely used assumption of having up to $3$ grants per subframe and let $W_\text{rar}$ be the number of consecutive subframes in a \gls{ra} response window.

The \emph{capacity} of the random access channels, that is, the maximum number of accessed UEs per unit of time, is limited by the period between RAOs $T_\text{rao}$, the number of available preambles at each RAO $R$, and the number of subframes for the \gls{ra} response window $W_\text{rar}$.


\section{Offloading: analysis and performance} \label{sec:offloading}

\subsection{Analysis}

Our analysis of the offloading scenario in  Figure~\ref{fig:backhauling_offloading}(a) focuses in the RA (to a terrestrial or a satellite \gls{gnb}) and the access latency performance when a satellite path is available. The possibility of using the satellite path implies that some or all devices are dual, i.e., they can choose at any time between connecting using the terrestrial or the space link. 

We first analyze the RA in the terrestrial path. An analogous analysis can be performed for the satellite path by changing the corresponding parameters. Let $R$ be the number of available preambles per RAO and $T_\text{rao}$ be the period between two consecutive RAOs in milliseconds. The terrestrial path receives a total rate $\lambda_\text{earth}$. 
 However, in a congested scenario with several supported services, it is likely that not all RAOs are available for a given service, as compared to the situation with low load and a single service. We model this with a larger value of terrestrial $T_\text{rao}$ (see Table~\ref{tab:parameter_settings}). 

Let $X_a$ be the random variable of the number of UEs that perform their $a$th access attempt at a given RAO, whose support is $x\in\{0,1,2,\dotsc,U\}$. Next, let $X=\sum_{a=1}^{A}X_a$ be the total number of contending UEs in the RAO. Therefore, $X_1$ is the number of UEs that initiate the RA in a RAO (i.e., those who transmit their first preamble). As described above, the number of newly generated updates is a Poisson process. Hence, the $X_1$ is a Poisson random variable with parameter $\lambda_\text{rao}=T_\text{rao}\lambda_\text{earth}$, whose pmf is 
\begin{equation}
    p_{X_1}(x)= \frac{\lambda_\text{rao}^x e^{-\lambda_\text{rao}}}{x!}
\end{equation}

Let $S$ be the RV of the number of preambles selected by a single UE in a specific RAO and $x$ be the number of contending UEs in the RAO. At any given RAO 
the expected value of $S$ for a given $x$ is
\begin{equation}
    \mathbb{E}\left[S; x\right]=x\left(1-\frac{1}{R}\right)^{x-1}\approx xe^{-\frac{x}{R}}
    \label{eq:expected_successes}
\end{equation}
From~\eqref{eq:expected_successes}, it is easy to observe that the number of successes is maximized when $x=R$. Hence, the maximum throughput in successful access attempts per second is
\begin{equation}
    \tau_\text{max}(R)=\frac{\max_x\mathbb{E}\left[S; x\right]}{T_\text{rao}}=\frac{R}{T_\text{rao}}\left(1-\frac{1}{R}\right)^{R-1}\approx \frac{R}{e\,T_\text{rao}}.
\end{equation}
Note that the latter corresponds to the maximum throughput of $R$ slotted ALOHA channels and indicates that the system is stable as long as $\mathbb{E}\left[X\right]\leq R/e$. That is, as long as the expected number of access attempts per RAO is lower than the maximum throughput. 

We say that a collision in a preamble occurs when this is selected by multiple UEs in the same RAO. Throughout this paper, we assume preamble collisions can be detected by the \gls{gnb} and, hence, the implicated \glspl{ue} do not receive a \gls{ra} response. From there, we define $C$ as the RV of the number of UEs that selected a preamble with collision. The expected value of $C$ for a given $x$ is given as
\begin{equation}
    \mathbb{E}\left[C\mid x\right]=x-x\left(1-\frac{1}{R}\right)^{x-1}=x\left(1-\left(1-\frac{1}{R}\right)^{x-1}\right)
\end{equation}

Hence, the probability of collision in a RAO with $x$ contending UEs is
\begin{equation}
    P_C(x)=1-\left(1-\frac{1}{R}\right)^{x-1}\approx 1-e^{-\frac{x}{R}}
\end{equation}
Conversely, the probability of selecting a unique preamble with $x$ contending UEs is
\begin{equation}
    P_S(x)=\left(1-\frac{1}{R}\right)^{x-1}\approx e^{-\frac{x}{R}}
    \label{eq:succ_prob}
\end{equation}

Next, we consider an erasure channel and define the erasure probability for a preamble transmission (i.e., failure due to a wireless channel error). Let $A$ be the maximum number of preamble transmission attempts that are allowed per each generated update. According to the recommendations of the 3GPP, the erasure probability for the $a$th preamble transmission for a specific update can be set to $\epsilon_a=1-e^{-a}$ for $a\in\{1,2,\dotsc,A \}$  to model the power ramping process~\cite{3GPPTR37868}. However, in delay- and \gls{aoi}-sensitive applications, a better strategy for the UEs is to use a relatively high power since the first transmission. Building on this, we assume a fixed erasure probability $\epsilon=\epsilon_a$ for all access attempts $a\in\{1,2, \dotsc,A\}$. Hence, a failed access attempt depends only on $x$ as follows
\begin{equation}
    P_f(x)= P_C(x) + (1-P_C(x))\epsilon
\end{equation}
Analogously, the probability of success for any value of $a$ is simply 
\begin{equation}
    P(x)= P_S(x)(1-\epsilon)
\end{equation}

All the UEs that fail an access attempt and that have not yet reached $A$ will backoff for a period selected randomly from $\left[0,T_\text{backoff}\right]$~milliseconds and transmit a newly selected preamble afterwards. Specifically, a UE that transmits its $a$-th preamble in a RAO where a total of $x$ preamble transmissions are performed will fail with probability $P_f(x)$. Then, if $a<A$, the UE will transmit a new preamble after the selected backoff period. On the other hand, if $a=A$, the UE declares a failure in the RA and concludes the process until the next update.

We then address the access delay, for which we consider the transmission of messages 1 to 4. 
Let $T_{access}(a)$ be the access delay of a UE that succeeds in the $a$th access attempt.
The minimum access delay is achieved when only one preamble transmission attempt is needed (i.e., with $a=1$, when no backoff is performed), the \gls{ra} response is sent at the beginning of the \gls{ra} window, and when no failures occur in the transmission of \emph{Msg3} and \emph{Msg4}. That is,
\begin{equation}
    \min {T}_{access} = t_\text{preamble} + t_{proc2} + t_\text{RAR}+t_{proc2}+t_{proc3}+t_{Msg3}+t_{Msg4}
\end{equation}

\noindent where
\begin{itemize}
    \item $t_\text{preamble}$ is the preamble duration;
    \item $t_\text{RAR}$ is the \gls{npdcch} RA response duration;
    \item $t_{Msg3}$ and $t_{Msg4}$ are the duration of \emph{Msg3} and \emph{Msg4}, respectively, set to $1$~ms;
    \item $t_{proc1}$, $t_{proc2}$, and $t_{proc3}$ are the preamble, uplink grant, and \emph{Msg3} processing times, respectively. These are assumed to be similar to those in earlier 3GPP releases $t_{proc1}=2$, $t_{proc2}=5$, and $t_{proc3}=4$~ms~\cite{3GPPTR36912}.
    
\end{itemize} 

Moreover, the RA response window defines the period in which \emph{Msg2} can be received~\cite{Leyva-Mayorga2017}. Remember that the number of RA responses per RAO is determined by the number of consecutive subframes in the RA response window and the maximum number of RA responses that can be sent per subframe in this window. 

Considering the duration of the preamble transmission, of the RA response window $W_\text{rar}$, and the backoff times, the latency when $a$ attempts are needed is given by
\begin{equation}
    {T}_{access}(a) =  \min T_{access} +t_\text{extra} + \sum_{v=1}^{a-1} \left(t_{BO}^{v}+ t_\text{preamble} + t_{proc1}+W_\text{rar}\right)
\end{equation}
\noindent where $t_{BO}^a$ is the backoff duration of attempt $a\in [1, A]$, drawn from a uniform distribution $U\sim [0,T_\text{backoff}]$ with $T_\text{backoff}$ the maximum backoff interval. Finally, $t_\text{extra}$ is the extra time spent queuing in the \gls{ra} response window and in the \gls{harq} processes during the transmission of \emph{Msg3} and \emph{Msg4}. A detailed model and analysis of the distribution of the access latency in LTE-A, with a similar access procedure, can be found in \cite{Leyva-Mayorga2017}.

In the terrestrial path, the preamble is repeated only once, therefore $\ell_\text{rep}=1$ and no additional modifications are needed to the 3GPP parameters and to the timing described above. In contrast, in the satellite path, the propagation delay has to be considered in the design of the protocol. For a \gls{leo} constellation at $600$ km, the minimum propagation delay is \mbox{$\approx 2$ ms} and the maximum with typical elevation angles is \mbox{$\approx 4$ ms}~\cite{Leyva-Mayorga2020}. 
To account for the difference in propagation delays, an extra prefix of $2$ ms is added at the beginning of the preamble transmissions. This compensates the difference in propagation delay between the worst and best cases in the coverage area of the satellite. 
Moreover, we consider preamble repetition, i.e. $\ell_\text{rep}>1$, to achieve an erasure probability comparable to the terrestrial path. This increases the effective \gls{sinr} at the \gls{gnb} and allows us to do a fair comparison of the performance of the two paths. Finally, the rapid movement of LEO satellites with respect to ground can lead to a large Doppler shift ~\cite{Leyva-Mayorga2020}. Given the narrowband nature of the \gls{nbiot} system (see Fig.~\ref{fig:ra_diagram}), mechanisms must be designed to achieve \gls{nbiot} connectivity between the ground and LEO satellites. However, it is out of the scope of this paper to design such mechanisms and, throughout the rest of the paper, we  assume that the Doppler shift is compensated at the space \glspl{gnb}. These aspects and other challenges for direct communication from the ground with space \glspl{gnb} are described in~\cite{3GPPTR38.821}. 


\subsection{Performance evaluation}
\begin{table}[t]
    \centering
     \caption{Parameter settings for the performance evaluation under the offloading and backhauling scenarios.}
    \begin{tabular}{@{}llll@{}}
    \toprule
    Parameter & Symbol & \multicolumn{2}{c}{Settings}\\\cmidrule{3-4}
    && Offloading & Backhauling\\
    \midrule
    \emph{Scenario}\\
    \enspace Total number of users & $U$ & $1000$ & $1000$\\
    \enspace Total update generation rate [updates per second]& $\lambda_\text{tot}$ &  $50$&$\{50,250\}$ \\
    \enspace Ground to total traffic ratio & $\kappa$ & $0.5$ & $1$\\
    \enspace Presence of a ground connection from the ground \gls{gnb} and the \gls{5gc} & $\mathbbm{1}_\text{5GC}$ & $1$ (True) & $0$ (False) \\
    \enspace Average service rate at the \gls{gnb} [updates per ms]& $\mu$ & $1$ & $1$\\
    \emph{Random access (RA)}\\
    \enspace Number of available preambles per \gls{rao} & $R$ &$36$ & $36$\\
     \enspace Number of repetitions (preamble and \gls{ra} response)& $\ell_\text{rep}$ & \{$1$ (ground) & $1$\\
      && \multicolumn{1}{r}{$4$ (space)\}}&\\
     \enspace Preamble duration [ms]& $t_\text{preamble}$ & $5.6\,\ell_\text{rep}$\\
    \enspace Period between \glspl{rao} for the ground \gls{gnb} [ms]& $ T_\text{rao}$ 
         & $\{160,320\}$ & $40$ \\
       \enspace Period between \glspl{rao} for the space \gls{gnb} [ms]& $T_\text{rao}^\text{space}$
         & $\{40,160\}$ & -- \\
    \enspace \Gls{ra} response window length [ms] & $W_\text{rar}$& $12\,\ell_\text{rep}$ & $12\,\ell_\text{rep}$\\
     \enspace Maximum number of \gls{ra} response messages in $W_\text{rar}$ & -- & $36$ & $36$\\
    \enspace Preamble duration per repetition [ms] & -- & $5.6$ & $5.6$\\
    \enspace Extended preamble prefix [ms] & -- & \{$0$ (ground),& $0$\\
    && \multicolumn{1}{r}{$2$ (space)\}}&\\
    \enspace Probability of erasure & $\epsilon$ & $0.1$ & $0.1$\\
    \enspace Maximum backoff period [ms] & $T_\text{backoff}$ & $\{160,320\}$ & $160$\\
    \enspace Maximum number of preamble transmission attempts per packet & $A$ & $\{1,10\}$& $\{0 \text{ (no RA)}, 1,10\}$\\
    \enspace Maximum propagation delay at the space \gls{gnb} [ms] & -- & $4$ & $0$\\
    \emph{Space segment}\\
    \enspace Number of satellites & $N$ &$1$ & $\{2, 4, 6\}$\\
    \enspace Probability of erasure at each satellite & $\varepsilon^s$ & $0$ & $\{0, 0.01, 0.1\}$\\
         \bottomrule
    \end{tabular}
  
    \label{tab:parameter_settings}
\end{table}

A summary of the parameters used in the performance analysis of the offloading and the backhauling cases is shown in Table~\ref{tab:parameter_settings}. 

\begin{figure}[t!]
	\centering
        \subfloat[]{\includegraphics{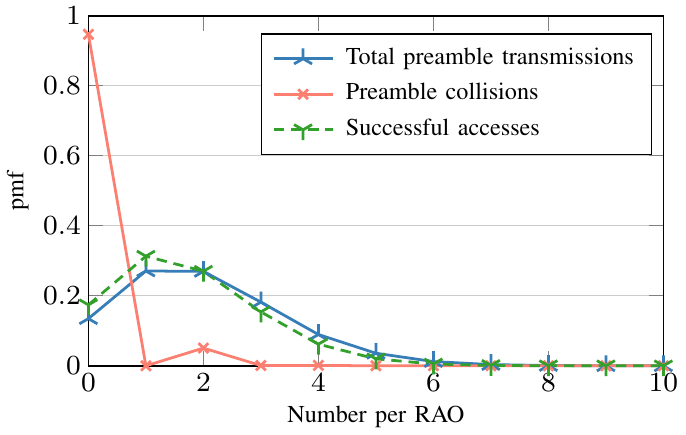}}\hfil
        \subfloat[]{\includegraphics{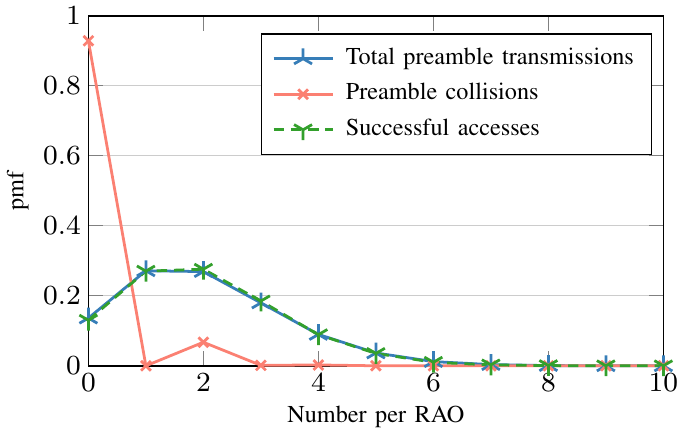}}
 \caption{\Gls{pmf} of the number of total, collided, and successful preamble transmissions per RAOs for (a) $A=1$  and (b) $A=10$ preamble transmission attempts under a light traffic load.}
 \label{fig:ul_backhaul}
\end{figure}

Figure~\ref{fig:ul_backhaul} plots the probability mass function of the \gls{ul} access with $A=1$ and $A=10$ preamble transmission attempts, valid for the terrestrial and satellite paths. With \mbox{1-attempt}, the probability of successful access is upper bounded by $(1-\epsilon)$, which corresponds to the case where a single \gls{ue} initiates the RA procedure per RAO. Conversely, performing more access attempts reduces the probability of failure due to errors in the transmission but also increases the load in the access channels. A similar behavior was previously observed for LTE-A~\cite{TelloOquendo2018}.

\begin{figure}
    \centering
    {\includegraphics{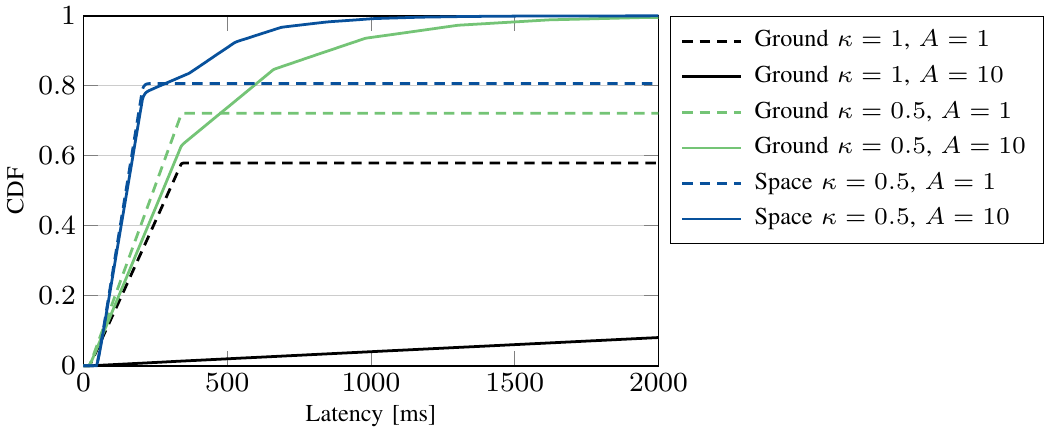}}
    \caption{Access latency for offloading. The period between \glspl{rao} in the ground \gls{gnb} is \mbox{$T_\text{rao}=320$~ms} and for the space \gls{gnb} is $T_\text{rao}^\text{space}=160$~ms with four preamble repetitions. The latency for failed accesses is set to $\infty$.}
    \label{fig:latency_offloading}
\end{figure}

Figure~\ref{fig:latency_offloading} shows the access latency in two cases: (1) all the traffic uses the terrestrial path, i.e., $\kappa = 1$; (2) the traffic is equally divided between the terrestrial and the satellite path, i.e., $\kappa = 0.5$. The latency for failed access is set to $\infty$, i.e., they are not included in the figure and therefore the CDF with only 1 attempt is upper bounded by $1-\epsilon$. In the ground path, $T_\text{rao} = 320$ ms, modeling a congested network that has to share the RAOs with other services. In the space \gls{gnb}, $T^\text{space}_\text{rao} = 160$ ms. It is observed that the addition of a satellite link can significantly alleviate the terrestrial network and improve the access latency, both with $A=1$ and $A=10$. The lowest latency is achieved by the spatial link. As expected, with $A=1$ the performance is relatively stable among the users and the CDF rapidly reaches the maximum value. Moreover, the performance with $A=1$ is also relatively stable among the distinct values of $\kappa$ (i.e., traffic loads) and across the space and terrestrial \glspl{gnb}. Conversely, the possibility of sending up to $10$ preambles highly increases the latency variance and becomes specially detrimental to the performance with $\kappa=1$. In this case, the multiple access attempts greatly increase the level of congestion in the \gls{nprach}, reaching a success probability of around $0.16$.

Notice that these results quantify the latency in the access network, but the differences in the end-to-end latency will also be impacted by the differences in the core network architecture and performance of the satellite path and the terrestrial path. This is left for future work. 


\section{Backhauling: analysis and performance} \label{sec:backhauling}


\subsection{Analysis}
The analysis of the backhaul scenario to connect a remote \gls{gnb} to the \gls{5gc} has two parts: (1) the random access from the device to the terrestrial \gls{gnb}; (2) the multi-hop space segment, as several \gls{leo} satellite are needed to relay the information if the \gls{gnb} and the \gls{5gc} are distant. For the RA, the analysis in Section~\ref{sec:offloading} is directly applicable, taking into account that $\kappa = 1$, i.e., all the traffic uses the terrestrial \gls{gnb}. Moreover, the \gls{gnb} is not shared with other services and therefore we use the value $T_\text{rao} = 40$ ms (see Table~\ref{tab:parameter_settings}). Therefore, we just include in this subsection the analysis of the latency and AoI in the space segment. 

\begin{figure}[t]
    \centering
    \includegraphics[width=6.5in]{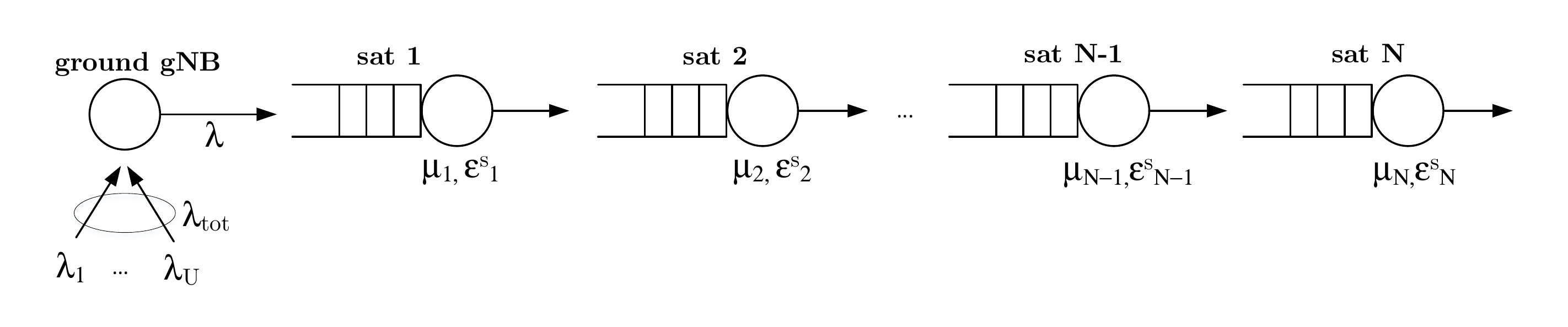}
    \caption{Queueing network model for backhauling } 
    \label{fig:queue_bh}
\end{figure}

The space segment consists of a multi-hop satellite system that relays the information from the terrestrial \gls{gnb}, connected to satellite $1$, to satellite $N$, using $N$ intermediate buffered-aided satellite nodes. The node $N$ is the closest satellite with a feeder link to connect to the 5GC. This relay network is modeled like a queueing network connected in series, as shown in Figure \ref{fig:queue_bh}, where each satellite buffer has infinite capacity and applies a \gls{fcfs} policy. It has been checked in the simulations that having a limited buffer has no big impact in the average latency and \gls{aoi} performance, unless the total buffer size is just few positions. 

In the queueing model of Figure~\ref{fig:queue_bh}, the first step is the link between the \gls{ue} and the \gls{gnb}, which uses RA. 
From the RA analysis we obtain the effective rate arriving to the first satellite $\lambda$. 
Then, from $n=1$ to $N-1$, the server model the \gls{isl} transmission between nodes. In the last node, $N$, the server models the feeder link to ground instead. The queueing network is characterized by the service rate vector $\bm{\mu}=(\mu_1,\ldots,\mu_N)$. The average service time of each satellite is the inverse of the service rate, $S_n = 1/\mu_n$. We model the $n$-th \gls{isl} between satellites $n$ and $n+1$ as a packet erasure channel with error probability $\varepsilon^s_n$. Thus, any packet sent by node $n$ is correctly received by node $n+1$ with probability $1-\varepsilon^s_n$. The overall Poisson arrival rate at node $n>1$ is then given by:
\begin{equation}
  \lambda_n= \lambda\prod_{\ell=1}^{n-1}(1-\varepsilon^s_\ell).\label{eq:lambda_err}
\end{equation}

Index $i$ is for packet $i$. Status update $i$ is generated at time $t_i$ and is received by the destination at time $t_i'$. Define $Y_i$ as the interarrival time $Y_i = t_i - t_{i-1}$ between two packets; $Z_i$ as the interdeparture time $Z_i = t'_i - t'_{i-1}$; and $T_i$ as the total network time in the system $T_i = t'_i - t_i$. The latter includes the time spent in all the nodes (queueing time and transmission time) until departure from the system at node $K$. 

For the analysis, we consider the service time in each node to be exponentially distributed. 
If all satellites apply the \gls{fcfs} queuing policy and the arrival process is Poisson, then the overall service time and waiting time follow a Hypoexponential distribution~\cite{amari1997closed}. The simplest case is when all departure rates are the same ($\mu = \mu_n\;\;\forall n$): in this case the total system time follows an Erlang distribution\footnote{The Hypoexponential distribution is indeed the convolution of several Erlang distributions.}, given by
\begin{equation}
f_T(t) = \frac{\alpha^N t^{N-1} e^{-\alpha t}}{(N-1)!} \label{eq:sojourn_distribution}
\end{equation}

However, due to the RA in the first node, the distribution is not Hypoexponential or Erlang, as the process at the first satellite is no longer Poisson. We consider the impact of the RA with two extreme cases: (1) For age-sensitive applications, a configuration with only one preamble transmission ($A = 1$) is suitable, as the failed sources prefer not to try again but just wait until the next status update is generated. In this case, the arrival process at the first satellite, after channel error and/or collision, is well approximated with a Poisson process for a wide range of arrival rates~\cite{Chiariotti2020}. (2) For a large value of $A$ the Poisson approximation is not valid and the RA has a significant contribution in the total latency and \gls{aoi}, as it will be confirmed in the evaluations with a large value $A=10$.  

If the system is in steady-state, the arrival rate at queue $n$ is the same as queue $n-1$, i.e., $\lambda_n = \lambda \;\;\forall n$. In this case and with equal service rates, we can conclude that the server utilization is also equal for all nodes and define $\rho_n = \rho = \frac{\lambda}{\mu} \;\; \forall n$. 
We define $\alpha = \mu (1-\rho)$ to be the parameter of the exponential distribution of each single M/M/1 stage, $f_{T_{M/M/1}}(t) = \alpha t e^{-\alpha t}$. Then, the average packet delay is given by the mean of the network delay distribution, i.e., \cite{Burke1956}
\begin{equation}
\bar{T} = \mathbb{E}[T] = \frac{N}{\alpha}  \label{eq:network_delay}
\end{equation}

\noindent where $\alpha=\mu-\lambda$

\begin{figure}
    \centering
    \subfloat{\includegraphics[width=3.5in]{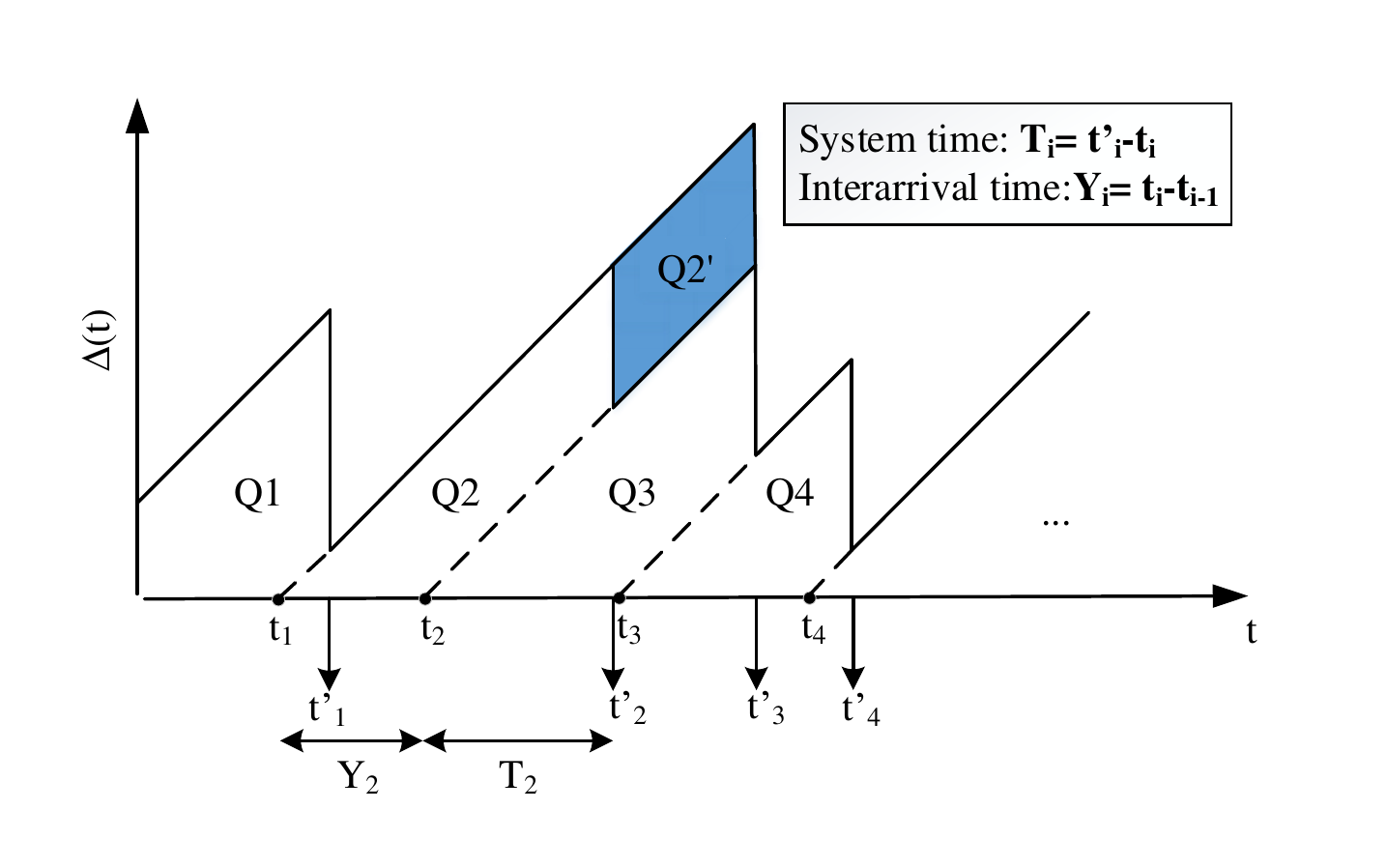}}\hfil
    \caption{Evolution of the Age of Information in a queue network with errors. The disjoint areas $Q_i$ are used in the derivation of the average AoI and related metrics \cite{Kosta2017}. The additional area $Q_2'$ highlighted in blue shows the increase in the AoI when one packet is lost. } 
    \label{fig:AoI}
\end{figure}

We can define the AoI in the destination at time $t$ as the random process $\Delta(t) = t - u(t)$. The age at the destination increases linearly in time in the absence of any updates, and is reset to a smaller value when an update is received. The evolution of the AoI $\Delta(t)$ at the destination exhibits the sawtooth pattern plotted in Figure \ref{fig:AoI}. Without loss of generality, the system is first observed at $t=0$ and the queue is empty with $\Delta(0) = \Delta_0$. 

The average AoI without errors is given as
\begin{equation}
    \bar{\Delta} = \frac{\mathbb{E}[YT] + \mathbb{E}[Y^2]/2}{\mathbb{E}[Y]}
\end{equation}

\noindent where ergodicity is assumed for the stochastic process $\Delta(t)$, $\mathbb{E}[Y] = 1/\lambda$ and $\mathbb{E}[Y^2]/2 = 1/\lambda^2$. 

If we consider the error rate $\varepsilon^s_n$ of each link, the expression must be rewritten to account for the extra trapezoid areas $Q_i'$ in Figure \ref{fig:AoI}.

After some calculations it yields~\cite{Chiariotti2020}

\begin{equation}
    \begin{aligned}
    \bar{\Delta}
            &=\frac{1}{\mathbb{E}[Y]} \sum_{e=0}^\infty p_s(N) \left(1-p_s(N)\right)^e\left(\mathbb{E}\left[T_i Y_i\right]+e\mathbb{E}\left[T_i Y_{i-1}\right]+\frac{e+1}{2}\mathbb{E}\left[Y_i^2\right]+\binom{e}{2}\mathbb{E}\left[Y_i\right]^2\right)
    \end{aligned}
\end{equation}

\noindent where $p_s(N) = \prod_{j=1}^N (1-\varepsilon^s_j)$.

The random variables $Y$ and $T$ are dependent and this complicates the calculations of the average age. In our model, the total system time of packet $i$ is the sum of the system times in each of the nodes $1, 2, ...N$, and each of them can be decomposed in waiting and service time
\begin{equation}
T_i = W_{i,1} + S_{i,1} + W_{i,2} + S_{i,2} + ... + W_{i,N} + S_{i,N}
\end{equation}

The expectation term can we decomposed
\begin{align}
\mathbb{E}[T_iY_{i}] &= \mathbb{E}[(W_i + S_i) Y_{i}] = \mathbb{E}[W_i Y_{i}] + \mathbb{E}[ S_i]E[Y_{i}] \nonumber \\ &= \mathbb{E}[W_i Y_{i}] + \sum_{n=1}^{N}\mathbb{E}[ S_{i,n}]\mathbb{E}[Y_{i}] = \mathbb{E}[W_i Y_i]  + \frac{N}{\mu \lambda} \label{eq:tiyi}
\end{align} 

The whole waiting time in the system is given by $W_{i,1}+...+W_{i,N}$. When packet $i$ arrives to the system there are two scenarios. Let us take the simplest case of two nodes. The first situation is that packet $i-1$ has already left the system when $i$ is generated, then $W_{i} = 0$. In the second possibility, packet $i-1$ leaves the system after the arrival of packet $i$. It can happen that when packet $i-1$ leaves the system, packet $i$ is in the first server; or it is waiting in the second queue. If packet $i$ does not find the last queue empty, we can write the total waiting time of packet $i$ in the general case of $K$ nodes as:  
\begin{equation}
W_{i} = \left(T_{i-1} - Y_{i} - \sum_{n=0}^{N-1}S_{i,n}\right)^+ = \left(T_{i-1} - Y_{i} - S_{\setminus N}\right)^+ 
\label{eq:wi}
\end{equation}

\noindent where we have defined $S_{\setminus N}=\sum_{n=1}^{N-1}S_{i,n}$ and $W_i$ is the total time spent in queues across the network, one or all of them, assuming a non-empty last queue. 

When the system reaches steady state the system times are stochastically identical, that is $T = ^{st} T_i =^{st} T_{i-1}$. 

Using the Erlang distribution for the system time and the Poisson arrival, and after some calculations, the expectation $\mathbb{E}[W Y]$ is tightly approximated by \cite{Soret2019}
\begin{align}
\mathbb{E}[W Y]  
&\approx \frac{-1}{\alpha \lambda^2 \Gamma(N)} \left( \alpha(\lambda S_{\setminus N}+2) \Gamma(N, \alpha S_{\setminus N}) - \lambda \Gamma(N+1, \alpha S_{\setminus N}) \right)  \nonumber
\\ 
&- \frac{ \alpha^N \mu^{-N} e^{\lambda S_{\setminus N}}}{\lambda^2 \mu \Gamma(N)}  
\cdot \left( \mu(\lambda S_{\setminus K}-2) \Gamma(N, \mu S_{\setminus N} ) - \lambda \Gamma(N+1, \mu S_{\setminus N}) \right);  \label{eq:I1}
\end{align}
\noindent where $\Gamma(s,x) = \int_x^{\infty}{t^{s-1}e^{-t} dt}$ and $\Gamma(s) = \int_0^{\infty}{t^{s-1}e^{-t} dt}$ are the upper incomplete and the ordinary gamma functions, respectively. 


\subsection{Performance evaluations}
\begin{figure}[t!]
	\centering
	\subfloat[]{
        \includegraphics{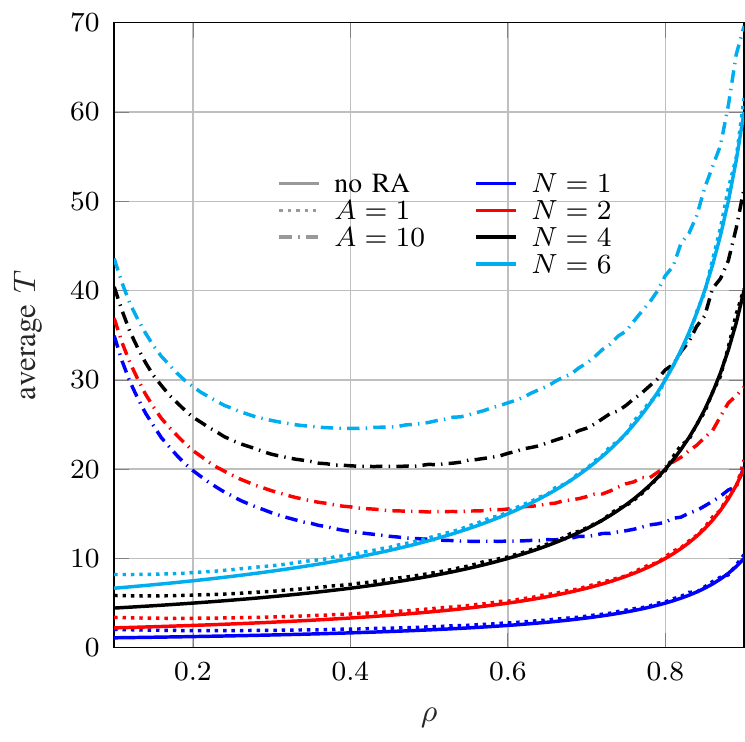}\vspace{-0.4cm}
        \label{fig:av_T}}\hfil
        \subfloat[]{\includegraphics{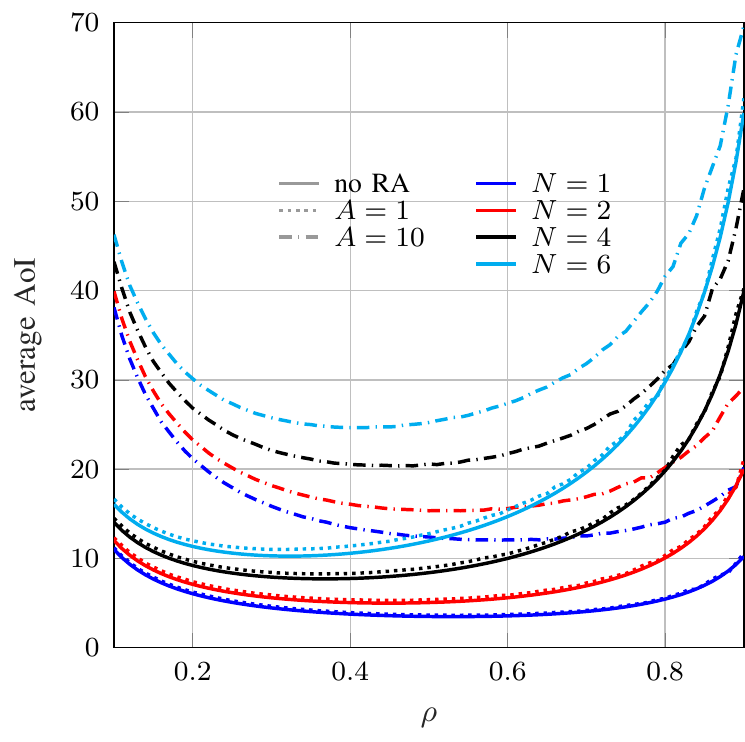}\vspace{-0.4cm}
        \label{fig:av_AoI}}
 \caption{Average system delay (left) and average AoI (right) of the backhaul system with no RA, $A=1$ and $A=10$ versus system load for an error-free satellite backhaul. The no RA case is modeled with a Poisson process as the arrival process in the first satellite. $N$ is the total number of satellites. $\mu_n=1\;\forall n$
 }
 \label{fig:av_aoi_T}
\end{figure}

\begin{figure}
	\centering
	\subfloat[]{
        \includegraphics{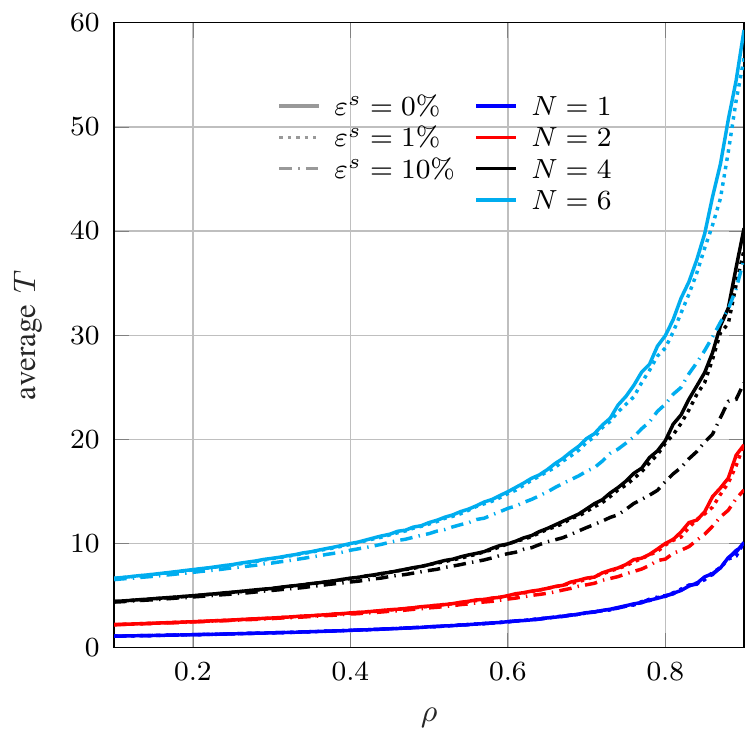}\vspace{-0.4cm}
        \label{fig:av_T_errors}}\hfil
	\subfloat[]{
        \includegraphics{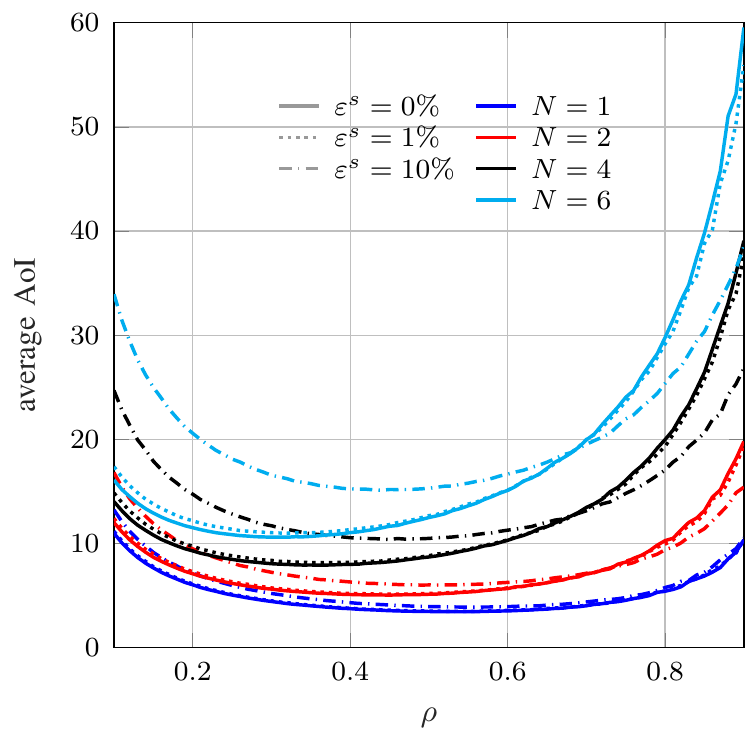}\vspace{-0.4cm}
        \label{fig:av_AoI_errors}}
 \caption{Average system delay (left) and average AoI (right) of the backhaul system versus system load for an error-prone satellite backhaul with different values of $\varepsilon^s$. $N$ is the total number of satellites. $\mu_n=1\;\forall n$
 }
 \label{fig:av_aoi_T_errors}
\end{figure}

Figure~\ref{fig:av_aoi_T} shows (a) the average system time and (b) the average AoI of the backhauling case versus the average rate at the first satellite ($\lambda= \rho$, as $\mu_n = 1\;\forall n$), for an error-free satellite network with $N=\{1,2,4,6\}$ satellites in the space segment. 
The two configurations of the RA access, with 1 and 10 attempts, are plotted. As expected, $\bar{T}$ and $\bar{\Delta}$ are always higher with 10 attempts. Moreover, having more satellites (more hops) is also detrimental for the latency behaviour. The solid line is the reference case where the users transmit their data directly to the satellite without performing the RA procedure (i.e., when the connection is maintained at all times). Note that this is a lower bound on latency and \gls{aoi}. 
The results for this cases are analytical, whereas the other two lines are obtained with Monte Carlo simulations of the queueing network and using as input to the first queueing system the departure process of the RA modeled in Section~\ref{sec:offloading}. Without RA or with 1-attempt RA, the average delay $\bar{T}$ increases with $\rho$, as expected. Moreover, the impact of having 1-attempt RA in the total $\bar{T}$ is minimal, which is visible in the small differences between the two cases. For 10-attempts RA, instead, $\bar{T}$ exhibits a very different behaviour with a U-shape and much larger values: the reason is that low values of load $\rho$ at the first satellite correspond to more collisions and therefore longer average delays before getting access, and this is the dominant component of the total delay. When the load increases the delay in the space segment becomes increasingly important and the three cases get closer for very high values of $\rho$. In the average AoI, the conventional U-shape is observed for the three cases. However, the results with 10-attempts RA separate from the other two for low values of $\rho$, where the uplink access is the dominant factor. When $\rho$ increases, the impact of the RA on the total average AoI is negligible and the results get closer to the solid line without RA. In this area, the high values of average AoI indicate that the frequency of the source packets is too high for the system capacity, which is working close to congestion and suffers from long queueing times. In both metrics, there is no specific unit in our results (i.e., they are given in symbols), as this will be given by the ISL frame and scheduling definition. 

Figure~\ref{fig:av_aoi_T_errors} plots the same two metrics, $\bar{T}$ and $\bar{\Delta}$, in an error-prone satellite backhaul (no RA). For simplicity, we set \mbox{$\varepsilon^s_n = \varepsilon^s = \{0, 0.01, 0.1\}$} so all the inter-satellite links experience the same error rate. 
The existence of errors is beneficial for the system delay performance, and this is more noticeable as the load increases, when it is more likely to find non-empty queues. In this situation, some packets queueing in the first nodes benefit from dropped packets in front because they experience less waiting time in the following queues. 
Interestingly, the packet losses are also good for the average age performance but only for high values of $\rho$. When the load is low, instead, the case without errors leads to the lowest average \gls{aoi}. 
This is because for low load the frequency of the source updates is the dominant factor, and queues are empty most of the time. Conversely, the impact of dropping packets and alleviating the queues becomes dominant for high $\rho$. 
Notice that the dropped packets are not included in the calculations of $\bar{T}$ and $\bar{\Delta}$.

\section{Conclusions} \label{sec:conclusions}
This paper studies the potential and performance of the access and the satellite segment in two relevant applications of a \gls{leo} constellation for 5G \gls{iot} networks: offloading, to alleviate a congested terrestrial network; and multi-hop backhauling, where there is no terrestrial cost-effective solution. In the offloading, it has been observed that having the possibility of using a satellite path alleviates a congested terrestrial network shared with other services, and the latency can still be improved for selected cases. The analysis of the random access reveals that performing more access attempts reduces the probability of failure due to errors in the transmission but also increases the load in the access channels. Having a single attempt (single preamble) is reasonable for some \gls{aoi} applications sending frequent updates and mostly interested in the freshess of the information at the receiver. For these cases, the results of system delay and \gls{aoi} with and without errors provide insights for the design of the backhaul topology and the number of required hops. 

There are several possible directions of extension of this work. First of all, the analysis of two architectural extensions already under discussion in 3GPP: multi-connectivity, with \gls{ue}s having the possibility to simultaneously transmit through both interfaces, terrestrial and satellite; and multi-link backhaul, where the traffic between the \gls{5gc} and the RAN is transmitted through one or more satellite links and one or more terrestrial links. Another important future direction is to consider the problems of the control plane and specifically the mobility of the satellites, as well as the characterization and evaluation of the latency and the throughput in the terrestrial \gls{5gc} segment, to have the full E2E performance figures. Finally, the models and performance evaluations are a valuable theoretical framework for future implementation and experimental work. 

\newpage

\bibliographystyle{IEEEtran}
\bibliography{5G_sat_terr_integ}

\end{document}